\documentclass[12pt]{article}
\def\be{\begin{equation}}
\def\ee{\end{equation}}
\def\bea{\begin{eqnarray}}
\def\eea{\end{eqnarray}}
\usepackage{graphicx}

\catcode`\@=11
\def\lsim{\mathrel{\mathpalette\@versim<}}
\def\gsim{\mathrel{\mathpalette\@versim>}}
\def\@versim#1#2{\vcenter{\offinterlineskip
\ialign{$\m@th#1\hfil##\hfil$\crcr#2\crcr\sim\crcr } }}
\catcode`\@=12
\usepackage{axodraw}

\catcode`@=12
\begin{document}
\thispagestyle{empty}
\begin{flushright}
UCRHEP-T472\\
August 2009\
\end{flushright}
\vspace{0.3in}
\begin{center}
{\LARGE \bf Neutrino Tribimaximal Mixing from A$_4$ Alone\\}
\vspace{1.5in}
{\bf Ernest Ma\\}
\vspace{0.2in}
{\sl Department of Physics and Astronomy, University of California,\\ 
Riverside, California 92521, USA\\}
\end{center}
\vspace{1.5in}
\begin{abstract}\
Neutrino tribimaximal mixing is obtained from the breaking of $A_4$ to 
$Z_3$ in the charged-lepton sector and to $Z_2$ in the neutrino sector.  
To enforce this conflicting pattern, extra particles and symmetries are 
usually invoked, often accompanied by nonrenormalizable interactions 
and even extra dimensions.  It is shown here in a specific 
renormalizable model how $A_4$ alone will accomplish this, with only the 
help of lepton number. 
\end{abstract}

\newpage
\baselineskip 24pt
\underline{\it Introduction}:~  The observed neutrino mixing matrix is very 
close to the tribimaximal form~\cite{hps02} and is best understood in terms 
of the tetrahedral symmetry $A_4$~\cite{mr01,bmv03,m04,af05}.  The key to 
its success is the pattern of symmetry breaking with preserved 
subgroups~\cite{l07,bhl08} such that $A_4 \to Z_3$ and $A_4 \to Z_2$ in 
two different sectors.  This misalignment is technically challenging to 
achieve and many auxiliary symmetries and particles have been invoked to 
make it happen~\cite{af05,bh05,af06,m06-1,h07,mpt07,m07,afh08,bfm08,m09,am09}, 
often with nonrenormalizable operators and even in the 
context of extra dimensions, thereby complicating the simple original idea 
of using just $A_4$.

On closer examination, it turns out that the breaking of $Z_3$ by $Z_2$ 
through a judicious choice of soft terms in the Higgs potential would 
work because there is a hidden symmetry which prevents the appearance of 
the other unwanted terms.  This is accomplished in a renormalizable 
model using Higgs doublets and triplets, both transforming as \underline{1} 
and \underline{3} of $A_4$ alone.

\underline{\it Lepton and Higgs Assignments}:~  The three families of 
leptons are considered~\cite{m06-2} as triplets under $A_4$:
\begin{equation}
L_i = (\nu_i,l_i) \sim \underline{3}, ~~~ l^c_i \sim \underline{3}.
\end{equation}
To obtain arbitrary lepton masses $m_{e,\mu,\tau}$, four Higgs doublets are 
used:
\begin{equation}
\Phi_0 = (\phi_0^0,\phi_0^-) \sim \underline{1}, ~~~ \Phi_i = 
(\phi_i^0,\phi_i^-) \sim \underline{3}.
\end{equation}
The Yukawa interactions are then given by
\begin{eqnarray}
{\cal L}_Y &=& h_0 (\nu_1 \phi_0^- - l_1 \phi_0^0) l_1^c +  h_1 (\nu_1 \phi_3^- 
- l_1 \phi_3^0) l_2^c +  h_2 (\nu_2 \phi_3^- - l_2 \phi_3^0) l_1^c \nonumber \\ 
&+& (1 \to 2, 2 \to 3, 3 \to 1) + (1 \to 3, 3 \to 2, 2 \to 1) + H.c.,
\end{eqnarray}
resulting in
\begin{equation}
{\cal M}_l = \pmatrix{h_0 v_0 & h_1 v_3 & h_2 v_2 \cr  h_2 v_3 & h_0 v_0 & 
h_1 v_1 \cr  h_1 v_2 & h_2 v_1 & h_0 v_0}.
\end{equation}
For $v_1=v_2=v_3=v$, this is exactly diagonalized:
\begin{equation}
{\cal M}_l = U_L \pmatrix{m_e & 0 & 0  \cr 0 & m_\mu & 0 \cr 0 & 0 & m_\tau} 
U_R^\dagger,
\end{equation}
where
\begin{equation}
U_L = U_R = {1 \over \sqrt{3}} \pmatrix{1 & 1 & 1 \cr 1 & \omega & \omega^2 
\cr 1 & \omega^2 & \omega},
\end{equation}
with $\omega = \exp(2 \pi i/3) = -1/2 + i\sqrt{3}/2$ and 
\begin{equation}
\pmatrix{m_e \cr m_\mu \cr m_\tau} = U_L \sqrt{3} \pmatrix{h_0 v_0 \cr h_1 v 
\cr h_2 v}.
\end{equation}

Before discussing the Higgs potential, note that the above is invariant 
under the transformation
\begin{eqnarray}
&& \Phi_1 \to \Phi_1, ~~ \Phi_{2,3} \to -\Phi_{2,3}, ~~~ v_1 \to v_1, ~~ 
v_{2,3} \to -v_{2,3}, \\ 
&& L_1 \to L_1, ~~ L_{2,3} \to -L_{2,3}, ~~~ l^c_1 \to l^c_1, ~~ l^c_{2,3} 
\to -l^c_{2,3}.
\end{eqnarray}
This hidden symmetry will prove to be crucial in obtaining tribimaximal 
mixing with $A_4$ alone.

\underline{\it Scalar Potential of Higgs Doublets}:~ Consider the scalar 
potential $V_0$ of $\Phi_0$ by itself.  It has just 2 terms:
\begin{equation}
V_0 = \mu_0^2 (\Phi_0^\dagger \Phi_0) + {1 \over 2} \lambda_0 (\Phi_0^\dagger 
\Phi_0)^2.
\end{equation}
The corresponding potential $V_1$ of $\Phi_{1,2,3}$ was shown to be~\cite{mr01}
\begin{eqnarray}
V_1 &=& \mu_1^2 \sum_i (\Phi_i^\dagger \Phi_i) + {1 \over 2} \lambda_1 
\sum_i (\Phi_i^\dagger \Phi_i)^2 + \lambda_2 |\Phi_1^\dagger \Phi_1 + \omega^2 
\Phi_2^\dagger \Phi_2 + \omega \Phi_3^\dagger \Phi_3|^2 \nonumber \\ 
&+& \lambda_3 \left[ |\Phi_2^\dagger \Phi_3|^2 + |\Phi_3^\dagger \Phi_1|^2 
+ |\Phi_1^\dagger \Phi_2|^2 \right] \nonumber \\ &+& \left\{ {1 \over 2} 
\lambda_4 \left[ 
(\Phi_2^\dagger \Phi_3)^2 + (\Phi_3^\dagger \Phi_1)^2 
+ (\Phi_1^\dagger \Phi_2)^2 \right] + H.c. \right\}.
\end{eqnarray}
The terms connecting the two are given by
\begin{eqnarray}
V_2 &=& \lambda_5 (\Phi_0^\dagger \Phi_0) \sum_i (\Phi_i^\dagger \Phi_i) + 
\left\{ \lambda_6 \sum_i (\Phi_0^\dagger \Phi_i)(\Phi_0^\dagger \Phi_i) + H.c. 
\right\} \nonumber \\ 
&+& \left[ \lambda_7 (\Phi_0^\dagger \Phi_1)(\Phi_2^\dagger \Phi_3) + 
(\Phi_0^\dagger \Phi_2)(\Phi_3^\dagger \Phi_1) + 
(\Phi_0^\dagger \Phi_3)(\Phi_1^\dagger \Phi_2) + H.c. \right] \nonumber \\ 
&+& \left[ \lambda_8 (\Phi_0^\dagger \Phi_1)(\Phi_3^\dagger \Phi_2) + 
(\Phi_0^\dagger \Phi_2)(\Phi_1^\dagger \Phi_3) + 
(\Phi_0^\dagger \Phi_3)(\Phi_2^\dagger \Phi_1) + H.c. \right].
\end{eqnarray}
It is clear from the above that the breaking of $A_4$ to $Z_3$ is maintained 
with $v_1=v_2=v_3$.  The new observation of here is that this solution 
is physically indistinguishable from $v_1 = -v_2 = -v_3$.

\underline{\it Neutrino Sector}:~ To obtain arbitrary Majorana neutrino 
masses, four Higgs triplets are used:
\begin{equation}
\xi_0 = (\xi_0^{++},\xi_0^+,\xi_0^0) \sim \underline{1}, ~~~ \xi_i = 
(\xi_i^{++},\xi_i^+,\xi_i^0) \sim \underline{3}.
\end{equation}
The Yukawa interactions are then of the form $\xi_0^0 \sum_i \nu_i \nu_i$ 
and $\xi_1^0 \nu_2 \nu_3 + \xi_2^0 \nu_3 \nu_1 + \xi_3^0 \nu_1 \nu_2$. 
It is now assumed that $u_2=u_3=0$, where $u_i = \langle \xi_i^0 \rangle$, 
so that $A_4$ breaks to $Z_2$, resulting in~\cite{m04,af05}
\begin{equation}
{\cal M}_\nu = \pmatrix{a & 0 & 0 \cr 0 & a & d \cr 0 & d & a} = U_\nu 
\pmatrix{a+d & 0 & 0 \cr 0 & a & 0 \cr 0 & 0 & -a+d} U_\nu^T,
\end{equation}
where
\begin{equation}
U_\nu = \pmatrix{1 & 0 & 0 \cr 0 & 1/\sqrt{2} & -1/\sqrt{2} \cr 0 & 
1/\sqrt{2} & 1/\sqrt{2}} \pmatrix{0 & 1 & 0 \cr 1 & 0 & 0 \cr 0 & 0 & i}.
\end{equation}
The mismatch between $U_L$ and $U_\nu$ yields~\cite{m04}
\begin{equation}
U_L^\dagger U_\nu = \pmatrix{\sqrt{2/3} & 1/\sqrt{3} & 0 \cr -1/\sqrt{6} & 
1/\sqrt{3} & -1/\sqrt{2} \cr -1/\sqrt{6} & 1/\sqrt{3} & 1/\sqrt{2}},
\end{equation}
i.e. tribimaximal mixing.  This is the simplest such realization, which 
is conistent with only the normal hierarchy of neutrino masses 
$(m_1 < m_2 < m_3)$, with the prediction~\cite{m05}
\begin{equation}
|m_{\nu_e}|^2 \simeq |m_{ee}|^2 + \Delta m^2_{atm}/9,
\end{equation}
where $m_{\nu_e}$ is the kinematical mass of $\nu_e$, and $m_{ee}$ is its mass 
measured in neutrinoless double beta decay.  Note that $\xi$ may be assigned 
the global lepton number $L=-2$ from its Yukawa couplings, which is then 
broken so that $L \to (-1)^L$.

\underline{\it Addition of Higgs Triplets}:~  It is assumed that $\xi_0$ 
and $\xi_{1,2,3}$ are all very heavy~\cite{ms98} with masses $M_0$ and $M_1$ 
respectively, so they do not appear as physical 
particles at or below the TeV scale.  They have quartic interactions among 
themselves similar to those of $V_{1,2,3}$.  It is easy to establish that 
the analog of Eq.~(8) holds here as well, i.e.
\begin{equation}
\xi_1 \to \xi_1, ~~~ \xi_{2,3} \to -\xi_{2,3}.
\end{equation}
To obtain the desirable solution $u_0 \neq 0$, $u_1 \neq 0$, and $u_2=u_3=0$, 
the global lepton number $L$ as well as $A_4$ must be broken.  The new 
observation of this note is that the following choice of soft scalar 
trilinear terms
\begin{equation}
V' = \mu'_0 \xi_0 \Phi_0 \Phi_0 + \mu'_1 \xi_0 \sum_i \Phi_i \Phi_i + 
\mu''_0 \xi_1 \Phi_0 \Phi_0 + \mu''_1 \xi_1 \sum_i \Phi_i \Phi_i
\end{equation}
works.  From $V'$, very small values of $u_0$ and $u_1$ are 
induced~\cite{ms98} by $v_0$ and $v_1 = v_2 = v_3 = v$, i.e.
\begin{equation}
u_0 \simeq {-\mu'_0 v_0^2 - 3 \mu'_1 v^2 \over M^2_0}, ~~~ 
u_1 \simeq {-\mu''_0 v_0^2 - 3 \mu''_1 v^2 \over M^2_1}, 
\end{equation}
but $u_{2,3}$ cannot appear.  To understand this, note first that 
in order for $\xi_0$ or $\xi_{1,2,3}$ to have a vacuum expectation value, $L$ 
must be broken and that can only be achieved through the terms of $V'$.  
However, $\xi_2$ or $\xi_3$ are only connected to these terms through 
their quartic or Yukawa couplings, which always preserve the symmetry 
$\psi_{2,3} \to -\psi_{2,3}$, where $\psi$ may be $(\nu,l)$ or $l^c$ or 
$\Phi$ or $\xi$.  Hence they always appear together and protect each 
other from getting a vacuum expectation value if neither has one 
to begin with.

\underline{\it Phenomenology of Higgs Doublets}:~  Since the Higgs triplets 
are very heavy, they may be integrated away, and from the explicit choice of 
$V'$, the effective scalar potential of Higgs doublets at the electroweak 
scale is still given by $V_0 + V_1 + V_2$.  Therefore, $A_4$ symmetry is 
maintained and its breaking $(v_1=v_2=v_3=v)$ results in a residual $Z_3$ 
symmetry, under which $\Phi_0$ and $\Phi'_0 = (\Phi_1 + \Phi_2 + \Phi_3)/
\sqrt{3}$ are singlets and
\begin{eqnarray}
\Phi'_1 = {1 \over \sqrt{3}} (\Phi_1 + \omega^2 \Phi_2 + \omega \Phi_3) \sim 
\omega, ~~~ 
\Phi'_2 = {1 \over \sqrt{3}} (\Phi_1 + \omega \Phi_2 + \omega^2 \Phi_3) \sim 
\omega^2.
\end{eqnarray}
The charged leptons are similarly redefined:
\begin{eqnarray}
e = {1 \over \sqrt{3}} (l_1 + l_2 + l_3) \sim 1, && 
e^c = {1 \over \sqrt{3}} (l^c_1 + l^c_2 + l^c_3) \sim 1, \\  
\mu = {1 \over \sqrt{3}} (l_1 + \omega l_2 + \omega^2 l_3) \sim \omega^2, && 
\mu^c = {1 \over \sqrt{3}} (l^c_1 + \omega^2 l^c_2 + \omega l^c_3) \sim \omega, 
\\  
\tau = {1 \over \sqrt{3}} (l_1 + \omega^2 l_2 + \omega l_3) \sim \omega, && 
\tau^c = {1 \over \sqrt{3}} (l^c_1 + \omega l^c_2 + \omega^2 l^c_3) \sim 
\omega^2.
\end{eqnarray}
As a result, $\phi_0^0$ couples to the charged leptons according to
\begin{equation}
{1 \over 3 v_0} (m_e + m_\mu + m_\tau) (e e^c + \mu \mu^c + \tau \tau^c),
\end{equation}
and ${\phi'}_0^0$ according to
\begin{eqnarray}
{1 \over 3 \sqrt{3} v} [(2m_e - m_\mu - m_\tau) e e^c + 
(2m_\mu - m_\tau - m_e) \mu \mu^c + (2m_\tau - m_e - m_\mu) \tau \tau^c] 
\end{eqnarray}
The linear combination $(v_0 \phi_0^0 + \sqrt{3} v {\phi'}_0^0)/\sqrt{v_0^2 + 
3v^2}$ corresponds to the Standard-Model Higgs boson which couples to 
$(m_e e e^c + m_\mu \mu \mu^c + m_\tau \tau \tau^c)/\sqrt{v_0^2+3v^2}$ as 
expected.

The Yukawa interactions of ${\phi'}_1^0$ and ${\phi'}_2^0$ are given by
\begin{eqnarray}
{1 \over 3 \sqrt{3} v} [
(2m_\mu - m_\tau - m_e) e \tau^c + 
(2m_\tau - m_e - m_\mu) \mu e^c + 
(2m_e - m_\mu - m_\tau) \tau \mu^c], \\ 
{1 \over 3 \sqrt{3} v} [
(2m_\tau - m_e - m_\mu) e \mu^c + 
(2m_e - m_\mu - m_\tau) \mu \tau^c + 
(2m_\mu - m_\tau - m_e) \tau e^c],
\end{eqnarray}
respectively.  Note that the above Yukawa couplings are different from 
those of the original $A_4$ model~\cite{mr01}, where $l^c_{1,2,3} \sim 
\underline{1}, \underline{1}', \underline{1}''$.  Note also that the 
residual $Z_3$ symmetry is a discrete lepton flavor symmetry which 
forbids~\cite{m09} transitions such as $\mu \to e \gamma$ and 
$\mu \to e e e$. However, there are processes forbidden by the usual 
lepton flavor $(L_e,L_\mu,L_\tau)$ conservation, but allowed by $Z_3$. 
From Eqs.~(27) and (28), it is clear that
\begin{equation}
\tau^+ \to e^+ e^+ \mu^-, ~~~ \tau^+ \to \mu^+ \mu^+ e^-, 
\end{equation}
are possible.  Their branching fractions are easily calculated to be
\begin{eqnarray}
B(\tau^+ \to e^+ e^+ \mu^-) &\simeq& B(\tau^+ \to \mu^+ \mu^+ e^-) \simeq 
\left( {2 m_\tau^2 \over 9 m_{eff}^2} \right)^2 \left( {v_0^2 + 3 v^2 \over 
3 v^2} \right)^2 B(\tau \to \mu \nu \nu) \nonumber \\ 
&\simeq& 8.6 \times 10^{-10} \left( {100~{\rm GeV} \over m_{eff}} \right)^4 
\left( {v_0^2 + 3 v^2 \over 3 v^2} \right)^2,
\end{eqnarray}
where $m_{eff}$ is the effective contribution of the two scalar fields 
${\phi'}_{1,2}^0$, as compared to the experimental upper bound of 
$1.1 \times 10^{-7}$.  Of course, if ${\phi'}_{1,2}^0$ are produced at 
the Large Hadron Collider (LHC), then their decays into two different 
charged leptons would be a unique signature of this model.

The contribution of ${\phi'}_{1,2}^0$ to the muon anomalous magnetic moment 
is given by
\begin{eqnarray}
\Delta a_\mu &=& {5 G_F m_\tau^2 \over 54 \sqrt{2} \pi^2} \left( {m_\mu^2 
\over m^2_{eff}} \right) \left( {v_0^2 + 3 v^2 \over 3 v^2} \right) \nonumber 
\\ &\simeq& 2.7 \times 10^{-13} \left( {100~{\rm GeV} \over m_{eff}} \right)^2 
\left( {v_0^2 + 3 v^2 \over 3 v^2} \right),
\end{eqnarray}
well below the current experimental sensitivity.

\underline{\it Conclusion}:~  It has been shown how a renormalizable model 
of Higgs doublets and triplets based on $A_4$ alone, with only the help of 
lepton number, is able to sustain neutrino tribimaximal mixing.  The key is 
the judicious choice of the soft terms of Eq.~(19), which break lepton 
number $L$ to $(-1)^L$, as well as $A_4$, but preserving the symmetry 
whereby all fields with indices 2 and 3 are reflected, i.e. $\psi_{2,3} 
\to -\psi_{2,3}$.  As a result, $\xi_{2,3}^0$ are protected by each other 
so that $u_{2,3}=0$ is a consistent solution, which then leads to 
tribimaximal mixing.  For the choice of Higgs doublets, there are also 
specific predictions of lepton flavor structure, which are testable 
at the LHC.

\noindent \underline{\it Acknowledgement}~:~ 
This work was supported in part by the U.~S.~Department of Energy under
Grant No. DE-FG03-94ER40837.

\baselineskip 16pt
\bibliographystyle{unsrt}

\end{document}